\documentstyle[12pt,aps,prd,preprint]{revtex}
\begin{document}
\title{Discrete Black-Hole Radiation and the Information Loss Paradox} 
\author{Shahar Hod}
\address{The Racah Institute of Physics, The
Hebrew University, Jerusalem 91904, Israel}
\address{and}
\address{Department of Condensed Matter Physics, Weizmann Institute, Rehovot 76100, Israel}
\date{\today}
\maketitle

\begin{abstract}

Hawking's black hole information puzzle highlights the incompatibility
between our present understanding of gravity and quantum
physics. However, during the last three decades evidence has been 
mounting that, in a quantum theory of gravity black holes may 
have a {\it discrete} line emission. 
A direct consequence of this intriguing prediction is that, 
black-hole radiance may carry a
significant amount of {\it information}. 
Using standard ideas from quantum information theory, we calculate the rate at
which information can be recovered from the black-hole spectral
lines. We conclude that the information that was suspected to be lost
may gradually leak back, encoded into the black-hole spectral lines.

\end{abstract}
\bigskip

Black-hole radiation, first predicted by Hawking \cite{Haw1}, imposes
a great challenge to our understanding of the interface between
quantum theory and gravity. As pointed out by Hawking, when a pure 
initial quantum state has collapsed to form a black hole, it will later evolve into a
high entropy mixed state of radiation. This contradicts one of the
basic principles of quantum mechanics -- unitary evolution, according
to which a pure state should always remain pure. This incompatibility 
is often discussed in terms of information theory
\cite{Beken1}: Since fully thermal radiation cannot convey detailed
information about its source, the information hidden in the black hole
about the state of the collapsed matter remains sequestered as the
black hole radiates, and it finally lost forever with the complete
evaporation of the black hole. This is the so-called black hole
information (loss) paradox.


Several different reactions to the black-hole 
information puzzle have been suggested, for reviews see e.g., \cite{Har,Gid1,Pre,Gid2}:
\begin{itemize}
\item {\it Information loss}. This point of view implies that the quantum theory of
  gravity inevitably violates the basic quantum mechanical principle of
  unitary evolution. Hawking \cite{Haw2} himself suggested a generalization of
  quantum mechanics which allows for an information loss, namely a one
  which permits the evolution of a pure state into a mixed
  state. However, specific schemes for such a generalization have been
  found to be incompatible with either locality of energy-momentum 
conservation \cite{Ell,Ban}.

\item {\it Remnants}. This point of view suggests that the black-hole
  evaporation stops as the black hole approaches the Planck scale
  (where Hawking's semi-classical analysis is expected to break down),
  and the remnant holds the information in question. The solution
  implies that there should be an infinite number of remnants species
  so that they can encompass the information associated with an
  arbitrarily large black hole \cite{Gid2}. But an infinite variety of
  remnant may imply an infinite production rates for remnants in
  processes like Hawking radiation, or Schwinger pair production (if
  the remnants are charged) \cite{Gid2}. Moreover, a Planck mass
  remnant can hold only a few bits of information \cite{Beken2,Gid3},
  much less than the entropy of a large evaporating black hole.

\item {\it Information leak}. Another alternative is that the 
  information which is supposed to be lost is encoded into the
  black-hole {\it radiation} and manages to leak back over the full duration
  of the evaporation process. In particular, Bekenstein \cite{Beken1}
  stressed the fact that Hawking's radiation departs from a blackbody
  radiation due to the mode dependence of the barrier penetration
  factor. This implies that Hawking's radiation is less entropic as
  compared with blackbody radiation (with the same power), and may
  therefore carry information with it. Bekenstein \cite{Beken1} 
has estimated that ({\it at least}) $1\% -5\%$ of the sequestered information
could come out in principle.
\end{itemize}

For information leak to be a physically reasonable resolution of
the puzzle one must show that information of the required 
magnitude can be encoded into the black-hole radiation. In the present
paper we put forward the idea that, when {\it quantum} properties of 
the {\it black hole itself} are properly taken into account, 
the information outflow rate may indeed be large enough
to allow a resolution of the paradox.

According to Hawking's result, the 
black hole emits quanta of {\it all} frequencies, distributed 
according to the usual black-body spectrum (with a gray-body factor 
which represents the imprint of passage through the curvature 
potential surrounding the black hole).

However, Hawking's prediction of black-hole evaporation is at a
semiclassical level in the sense that the matter fields are treated
quantum mechanically, but the spacetime (and the black hole itself)
are treated {\it classically}. One therefore suspects some
modifications of the character of the radiation when quantum
properties of the {\it black hole itself} are properly taken into account.

The quantization of black holes was proposed long ago by Bekenstein
\cite{Beken3}. Based on the remarkable observation that the horizon
area of a nonextremal black hole behaves as a classical 
adiabatic invariant, and in the spirit of Ehrenfest principle
\cite{Ehren}, any classical adiabatic invariant
corresponds to a quantum entity with {\it discrete} spectrum,
Bekenstein \cite{Beken3} conjectured that the horizon area of a quantum
black hole should have a discrete eigenvalue spectrum of the form

\begin{equation}\label{Eq1}
A_n=\gamma {\ell^2_P} \cdot n\ \ \ ;\ \ \ n=1,2,\ldots\ \  ,
\end{equation}
where $\gamma$ is a dimensionless constant, and 
$\ell_P=\left({G \over {c^3}}\right)^{1/2} {\hbar}^{1/2}$ is the
Planck length (we use gravitational units in which $G=c=1$). This type
of quantization-law has since been revived on various grounds
\cite{Muk,BekMuk,Beken4,Maz,Ko,DanSch,Mag,Lo,Pe,LoMa,BaKu,Ka1,MakRep,Hod1,Hod2,Hod3,VaWi,Ka2,BoKa,Ah,Ga1,Ga2}
(most of these derivations have been made in the last few years). 
In particular, Mukhanov and Bekenstein
\cite{Muk,BekMuk,Beken4} used a combination of thermodynamic 
(the area-entropy relation $S_{BH}=A/4\hbar$ for
black holes) and statistical physics (the Boltzmann-Einstein
formula) arguments, to find that the dimensionless constant $\gamma$ in
Eq. (\ref{Eq1}) should be of the form $\gamma =4 \ln \beta$, with $\beta=2,
3, \ldots$ (this corresponds to a degeneracy factor of $\beta^n$ for the
$n$th area level). Using Bohr's correspondence principle Hod \cite{Hod1}
has recently given evidence in favor of the value $\beta=3$.

The discrete mass (area) spectrum implies
a {\it discrete} line emission from a quantum black hole; the
radiation emitted by the black hole will be at integer multiples of
the fundamental frequency $\omega_0=\ln \beta/8\pi M$ \cite{BekMuk}. 
The broadness of the lines will be discussed at the end of the paper. 

We note that a direct consequence of the {\it discrete} spectrum is
that, compared with blackbody radiation (or even with Hawking's
semi-classical continues radiation), black-hole 
radiance is {\it less} entropic. The entropic deficiency may permit 
an information outflow of the required magnitude 
(In fact, we shall show below that the rate of information outflow {\it increases} as the
spacing between the discrete energy levels increases.) 
Thus, information 
about the quantum state of the collapsed matter 
may in principle be encoded into the discrete black-hole spectral
lines. From the point of view of quantum communication theory (for
reviews see \cite{Yam,BekSch}), the
maximum rate at which information can be recovered from the radiation
is $\dot I_{max} \equiv \dot S^{'}-\dot S$, where $\dot S$ is the
{\it actual} entropy outflow rate and $\dot S^{'}$ is 
the {\it maximum} rate for entropy outflow corresponding to 
the {\it actual} power under the boundary conditions of the physical
system \cite{Beken1}.

The probability for a black hole to emit a specific quantum 
should be proportional to the 
degeneracy of the {\it final} black-hole quantum state (and thus should be 
proportional to $\beta^{-k}$, $k$ being the level spacing between the initial and final
quantum states), to the gray-body
factor $\Gamma$ (representing a scattering of the quantum 
off the spacetime curvature surrounding the black hole), and to
the square of the matrix element. In the spirit of the original
treatment of Bekenstein and Mukhanov \cite{BekMuk} 
we assume that the matrix element does not vary much as
one goes from a nearest neighbor transition to one between somewhat
farther neighbors. Thus, aside from an overall normalization 
factor, the matrix element does not enter into our 
simple estimate. This assumption is further supported by a recent
analysis of Massar and Parentani \cite{MaPa}. Thus, the probability
$p_k$ to jump $k$ steps in the mass (area) ladder is proportional to
$\Gamma(k) \beta^{-k}$. 

We consider the emission of a canonical set of three
species of neutrinos, photon, and a graviton. This is the 
set of particles considered in former analyses of black-hole
evaporation \cite{Page,Beken1,Cham}. (The physical
implications of a massive neutrino field were discussed in
\cite{Hod4}.) For the power emitted by the quantized black hole we 
write $\dot E=\epsilon \hbar \omega_0 /\tau$, where $\epsilon \hbar \omega_0$ is the
mean energy carried away by an emitted quanta, and $\tau$ is the mean time
between quantum leaps. Later on we shall estimate a lower 
bound on $\tau$. The coefficient 
$\epsilon$ depends on
the gray-body factors and should therefore be evaluated
numerically. One finds $\epsilon \omega_0=0.185M^{-1}$ for 
$\beta=2$, with a deviation of less than $1\%$ for $\beta=3$. 
This last result also implies that the mean
decrease in black-hole entropy 
with each quanta emitted, which is given by $\epsilon \ln \beta$, 
is $4.65$ for $\beta=2$ (again, with a deviation of less than $1\%$
for $\beta=3$). [Each quanta emitted from the quantum black hole decreases
its mass by $\Delta M=-\epsilon \hbar \ln \beta/8\pi M$ on the average. 
Using the relation $S_{BH}=4\pi M^2/ \hbar$, one 
finds that the mean decrease in black-hole entropy with each quanta emitted is $\epsilon \ln \beta$, 
or $\epsilon \omega_0 8\pi M =4.65$.]

The entropy of a system measures one's {\it lack of information} about
its actual internal configuration \cite{Shan,Jayn,Beken5}. 
Suppose that all that is known about the system's internal configuration is
that it may be found in any of a number of states, with probability
$p_n$ for the $n$th state. Then the entropy associated with the system
is given by Shannon's well-known relation $S=-\sum_{n} p_n \ln p_n$. 
The ratio $R=|\dot S_{rad} / \dot S_{BH}|$ of entropy emission
rate from the quantum black hole, to the rate of black-hole
entropy decrease is therefore given by
             
\begin{equation}\label{Eq2}
R={{-{\sum_{i=1}^{N_s}}{\sum_{k=1}^{\infty}}C\Gamma(k)\beta^{-k} \ln
  [C\Gamma(k)\beta^{-k}]} \over {{\sum_{i=1}^{N_s}}{\sum_{k=1
}^{\infty}}C \Gamma(k)\beta^{-k} k \ln\beta}}\  ,
\end{equation}
where $N_s$ is the effective number
of (massless) particle species emitted ($N_s$ takes into account the
various modes emitted), and $C$ is a normalization factor [Note that the 
denominator in Eq. (\ref{Eq2}) is simply $\epsilon \ln \beta$.] 
The ratio $R$ was calculated in \cite{Hod4}: $R=1.119$ for $\beta=2$
and $R=1.016$ for $\beta=3$. 
Thus, the mean entropy carried away by an emitted quanta is 
$4.65R$, and the corresponding rate of entropy emission 
from the quantum black hole is given by 

\begin{equation}\label{Eq3}
\dot S_{rad} =4.65R/\tau\  .
\end{equation}

One now has to compare $\dot S_{rad}$ with the maximally entropic 
(blackbody) distribution $\dot S^{'}$ whose power $\dot E^{'}$ equals the
actual power $\dot E$ of the black-hole radiation. According to the
Boltzmann formula, a black body at temperature $T$ and of area $A$ emits 
power $\dot E^{'}=N\pi^2 A T^4/60\hbar^3$, where $N$ is the
effective number of particle species emitted. Photons and gravitons
contribute $1$ each to $N$, while a neutrino contributes $7/16$. Thus,
we should take $N=37/8$ for the canonical set. Now,
for blackbody radiation flowing in three space dimensions, 
$\dot S^{'}={4 \over 3}\dot E^{'}/T$. Thus, one has the relation 

\begin{equation}\label{Eq4}
\dot S^{'}={{4 N^{1/4}\pi^{1/2}A^{1/4}{\dot E^{'3/4}}} \over { 4860^{1/4}\hbar^{3/4}}}\  .
\end{equation}
The effective radiating area of a black hole (the ``photosphere'')
was estimated by Bekenstein \cite{Beken1} as $A_{phot}=108 \pi M^2
\xi$, with $\xi >1$. This is a reasonable estimation because in the
high frequency regime, all quanta hitting within a cross section $27
\pi M^2$ will be captured \cite{MTW,Wal} (this suggests a photospheric
area four times larger) . The factor $\xi$ takes into
account the fact that the gray-body factors $\Gamma(\omega)$ vanish
only as a power-law of $\omega$ in the $\omega \to 0$ limit, and thus
quanta of a fairly {\it large} impact parameter are sometimes absorbed by the
black hole, and therefore may be emitted sometimes \cite{Beken1}. 
Substitution of the numerical factors in Eq. (\ref{Eq4}) yields

\begin{equation}\label{Eq5}
\dot S^{'}=1.507 \xi^{1/4}M^{-1/4} \tau^{-3/4}\  . 
\end{equation}

Taking cognizance of Eqs. (\ref{Eq3}) and (\ref{Eq5}) one finally
finds that the maximum rate at which
information can leak out from the quantum black hole (carried away by
the black-hole spectral lines) is

\begin{equation}\label{Eq6}
\dot I_{max} \equiv \dot S^{'}-\dot S_{rad}=(1.507 \xi^{1/4}\alpha^{1/4}-
4.65R)/\tau\  ,
\end{equation}
where $\alpha \equiv \tau/ M$.

If the rate at which {\it information} leaks out from the black
hole amounts to $4.65R/\tau$ (which is the actual {\it entropy} emission 
rate), than given an appropriate quantum
mechanism this information may reduce one's uncertainty 
(lack of information) about the actual internal configuration of the
system, in accord with the general formula $\Delta I=-\Delta S$ 
relating information and entropy \cite{Shan,Jayn,Beken5}. 
Thus, the radiation can end up in a pure
state. We should therefore evaluate the ratio 
$T \equiv \dot I / \dot S_{rad}$, which is given by

\begin{equation}\label{Eq7}
T \equiv  \dot I_{max} / \dot S_{rad} =
\left\{
 \begin{array}{l@{\quad,\quad}l}
0.289\xi^{1/4}\alpha^{1/4} - 1 & \beta=2  \\
0.319\xi^{1/4}\alpha^{1/4} - 1 & \beta=3  
\end{array} \right. \    
\end{equation}
(Here we used the values of $R$ calculated in \cite{Hod4}.)

We shall now estimate a lower bound on 
$\alpha$ (which would yield a lower bound on the
physically interesting quantity $T$). To that end 
we use Page's \cite{Page} semi-classical 
result for the emitted power and write $\dot E=2.829 \cdot 10^{-4} \eta \hbar/M^2$. 
The factor $\eta$ takes into account the fact that 
the emission is in {\it discrete} lines, and thus part of the 
frequency spectrum is blocked. This should suppress the power emitted 
from the {\it quantized} black hole as compared with the 
semi-classical continuous spectrum. One therefore expects the
coefficient $\eta$ to be smaller than unity. On the other hand, we have already
found that the mean energy carried by a quanta emitted from the
quantum black hole is $0.185 \hbar M^{-1}$, and the emitted power therefore 
equals $\dot E =0.185 \hbar M^{-1}/\tau$. Comparing the two expressions
one finds $\tau \simeq 654 \eta^{-1}M$, which 
implies $\alpha \simeq 654 \eta^{-1}$. The corresponding values of $T$
are given in Table \ref{Tab1}, from which we learn that the
information outflow rate may be of just the required 
magnitude ($T \geq 1$) if $\sim 50\% - 70\%$ of the power (or more) 
is blocked \cite{note1}.

Note that the value of $T$ derived in this paper (for a {\it quantum} 
black hole) is higher than the 
corresponding one given in \cite{Beken1} (which was derived for a 
{\it classical} black hole). 
In other words, the radiation emitted from a quantum black 
hole carries with it more information than the one coming out of a classical 
black hole. This result is a direct consequence of the fact that the 
spectral emission from a quantum black hole is in {\it discrete} lines, 
whereas a classical black hole emits radiation in the form of 
a smooth {\it continuum}. This implies a larger deviation from a 
perfect (maximally entropic) blackbody radiation in the case of a 
quantum black hole, as compared with a classical one.

For our analysis to be self-consistent, the natural broading of the
spectral lines should {\it not} smear the
spectrum into a smooth continuum. The question of
natural broading of the black-hole (discrete) spectral lines has been
discussed previously by Bekenstein and Mukhanov \cite{BekMuk} and 
by  M\"akel\"a \cite{Mak}. These analyses are, however,
semi-qualitative and in particular do not take fully into account the central role of
the gray-body factors. Here we take these into account. 
We recall that,  according to Heisenberg's quantum uncertainty
principle the natural broading
$\delta \omega$ is of the order of $\tau^{-1}$, which yields the ratio

\begin{equation}\label{Eq8}
{{\delta \omega} \over {\omega_o}} \sim 0.035 \eta\  ,
\end{equation}
for $\beta=3$, and $\delta \omega / \omega_0 \sim 0.055 \eta$ for
$\beta=2$. This implies that the emission lines are unblended, in agreement with
the semi-qualitative estimation of Bekenstein and Mukhanov
\cite{BekMuk} (However, the numerical ratio $\delta \omega / \omega_0$
is different because in
\cite{BekMuk} only one emitted specie was considered, and the gray-body
factors were not taken into account.)

In summary, we have shown that if {\it quantum} properties
of the {\it black hole} itself are properly taken into account, 
the information which was suspected to be lost may in fact be 
encoded into the black-hole spectral lines. Thus, with the help of 
an appropriate quantum mechanism to encode the information, 
the reconstruction of a pure radiation state seems physically
reasonable. For the advocates of the ``information leak'' resolution the
task remains to identify the appropriate quantum mechanism.

\bigskip
\noindent
{\bf ACKNOWLEDGMENTS}
\bigskip

I thank Jacob D. Bekenstein for helpful discussions.
This research was supported by grant 159/99-3 from the Israel 
Science Foundation.

\begin{table}
\caption{The ratio $T \equiv \dot I / \dot S$ of the maximal {\it information} 
outflow rate to the actual {\it entropy} emission rate, and the natural
broading $\delta \omega / \omega_0$ of the spectral lines.}

\label{Tab1}
\begin{tabular}{lcl}
$\beta$ &$T$ & $\delta \omega / \omega_0$\\
\tableline
$2$ & $1.461 \xi^{1/4} \eta^{-1/4} - 1$ & $0.055 \eta$ \\
$3$ & $1.613 \xi^{1/4} \eta^{-1/4} - 1$ & $0.035 \eta$ \\
\end{tabular}
\end{table}


\begin{thebibliography}{99}

\bibitem{Haw1} S. W. Hawking, Commun. Math. Phys. {\bf 43}, 199 (1975).

\bibitem{Beken1} J. D. Bekenstein, Phys. Rev. Lett. {\bf 70}, 3680 (1993).

\bibitem{Har} J. Harvey and A. Strominger, ``Quantum Aspects of Black
  Holes'', e-print hep-th/9209055.

\bibitem{Gid1} S. B. Giddings, ``Toy Models for Black Hole Evaporation'',
  e-print hep-th/9209113.

\bibitem{Pre} J. Preskill, ``Do Black Holes Destropy Information'',
  e-print hep-th/9209058.

\bibitem{Gid2} S. B. Giddings, ``Introduction to the Information
  Problem'', e-print astro-ph/9412046.

\bibitem{Haw2} S. W. Hawking, Commun. Math. Phys. {\bf 87}, 395 (1982).

\bibitem{Ell} J. Ellis, J. Hagelin, D.V. Nanopoulos and M. Srednicki,
  Nucl. Phys. B {\bf 241}, 381 (1984).

\bibitem{Ban} T. Banks, M. E. Peskin and L. Susskind,  Nucl. Phys. B {\bf 244}, 125 (1984).

\bibitem{Beken2} J. D. Bekenstein, Phys. Rev. D {\bf 23}, 287 (1981).

\bibitem{Gid3} S. B. Giddings, Phys. Rev. D {\bf 46}, 1347 (1992).

\bibitem{Beken3} J. D. Bekenstein, Lett. Nuovo Cimento {\bf 11}, 467 (1974).

\bibitem{Ehren} See for example M. Born, Atomic Physics (Blackie,
  London, 1969), eighth edition.

\bibitem{Muk} V. Mukhanov, JETP Lett. {\bf 44}, 63 (1986).

\bibitem{BekMuk} J. D. Bekenstein and V. F. Mukhanov, Phys. Lett. B
  {\bf 360}, 7 (1995).

\bibitem{Beken4} J. D. Bekenstein in {\it Proceedings of the XVII Brazilian National Meeting on Particles and Fields}, edited by
 A. J. da Silva et. al. (Brazilian
  Physical Society, Sao Paulo, 1996); 
J. D. Bekenstein in {\it Proceedings of the VIII Marcel Grossmann Meeting on General Relativity}, edited by T. Piran and
  R. Ruffini (World Scientific , Singapore, 1998).

\bibitem{Maz} P. O. Mazur, Phys. Rev. Lett. {\bf 57}, 929 (1987).

\bibitem{Ko} Ya. I. Kogan, JETP Lett. {\bf 44}, 267 (1986).

\bibitem{DanSch} U. H. Danielsson and M. Schiffer, Phys. Rev. D {\bf 48}, 4779 (1993).

\bibitem{Mag} M. Maggiore, Nucl. Phys. B {\bf 429}, 205 (1994).

\bibitem{Lo} C. O. Lousto, Phys. Rev. D {\bf 51}, 1733 (1995).

\bibitem{Pe} Y. Peleg, Phys. Lett. B {\bf 356}, 462 (1995).

\bibitem{LoMa} J. Louko and J. M\"akel\"a, Phys. Rev. D {\bf
    54}, 4982 (1996).

\bibitem{BaKu} A. Barvinsky and G. Kunstatter, Phys. Lett. B {\bf
    389}, 231 (1996).

\bibitem{Ka1} H. A. Kastrup, Phys. Lett. B {\bf 385}, 75 (1996).

\bibitem{MakRep} J. M\"akel\"a and P. Repo, Phys. Rev. D {\bf 57}, 4899 (1998).

\bibitem{Hod1} S. Hod, Phys. Rev. Lett. {\bf 81}, 4293 (1998).

\bibitem{Hod2} S. Hod, Phys. Rev. D {\bf 59}, 024014 (1999).

\bibitem{Hod3} S. Hod, Gen. Rel. Grav. {\bf 31}, 1639 (1999).

\bibitem{VaWi} C. Vaz and L. Witten, Phys.Rev. D {\bf 60}, 024009 (1999).

\bibitem{Ka2} H. A. Kastrup, Annalen Phys. {\bf 9}, 503 (2000).

\bibitem{BoKa} M. Bojowald and H. A. Kastrup, Class.Quant.Grav. {\bf 17}, 3009 (2000).

\bibitem{Ah} D. V. Ahluwalia, Int. J. Mod. Phys. D {\bf 8}, 651 (1999).

\bibitem{Ga1} R. Garattini, Nucl.Phys.Proc.Suppl. {\bf 88}, 297 (2000).

\bibitem{Ga2} R. Garattini, e-print gr-qc/0003090.

\bibitem{Yam} Y. Yamamoto and H. A. Haus, Revs. Mod. Phys. {\bf 58},
  1001 (1986).

\bibitem{BekSch} J. D. Bekenstein and M. Schiffer,
  Int. Journ. Mod. Phys. C {\bf 1}, 355 (1990).

\bibitem{MaPa} S. Massar and R. Parentani, Nucl.Phys. B {\bf 575}, 333 (2000).

%


\bibitem{Page} D. N. Page, Phys. Rev. D {\bf 13}, 198 (1976);
  Phys. Rev. D {\bf 14}, 3260 (1976); Phys. Rev. D {\bf 16}, 2402 (1977).

\bibitem{Cham} C. M. Chambers, W. A. Hiscock and  B. Taylor,
  Phys. Rev. Lett. {\bf 78}, 3249 (1997).

\bibitem{Hod4} S. Hod, Phys. Rev. D {\bf 61}, 124016 (2000).

\bibitem{Shan} C. E. Shannon and W. Weaver, 
{\it The Mathematical Theory of Communications} (University of Illinois Press, Urbana, 1949).

\bibitem{Jayn} E. T. Jaynes, Phys. Rev. {\bf 106}, 620 (1957); {\bf 108}, 171 (1957).

\bibitem{Beken5} J. D. Bekenstein, Phys. Rev. D {\bf 7}, 2333 (1973).

\bibitem{MTW} C. M. Misner, K. S. Thorne and J. A. Wheeler, {\it Gravitation} (Freeman, San Francisco, 1973).

\bibitem{Wal} R. M. Wald, in {\it Quantum Field 
Theory in Curved Spacetime and Black Hole Thermodynamics} (The University of Chicago
    Press, Chicage 1994).

\bibitem{note1} Note that, as previously conjectured, the rate of
  information outflow {\it increases} as the
spacing between the (discrete) energy levels increases (see Table
\ref{Tab1}). This is caused by the
fact that the entropy of the
radiation should be maximal when the various transitions have equal
probabilities, but the fundamental transition $n \to n-1$ becomes more
and more dominant as the value of $\beta$ increases. Thus, $\dot S$ is
a decreasing function of $\beta$, which implies that $\dot I_{max}$
increases with $\beta$.

\bibitem{Mak} J. M\"akel\"a, Phys. Lett. B {\bf 390}, 115 (1997).


\end{thebibliography}
\end{document}